\documentclass[twocolumn,showpacs,preprintnumbers,superscriptaddress,amsmath,amssymb,floatfix,prl]{revtex4}
\usepackage{graphicx}

\def\nm{{\ {\rm nm}}}						
\def\mm{{\ {\rm mm}}}						
\def\cm{{\ {\rm cm}}}						
\def\micron{{\ \mu{\rm m}}}					

\def\mms{{\ {\rm mm}/{\rm s}}}				

\def\Hz{{\ {\rm Hz}}}						
\def\kHz{{\ {\rm kHz}}}						
\def\GHz{{\ {\rm GHz}}}						

\def\us{{\ \mu{\rm s}}}						
\def\ms{{\ {\rm ms}}}						

\def\Schrodinger{{Schr\"odinger\ }}

\def\Er{{{E_R}}}							
\def\kr{{{k_R}}}							
\def\Rb87{^{87}\rm{Rb}}					

\begin{document}

\title{Collisional de-excitation in a quasi-2D degenerate Bose gas}

\author{I.~B.~Spielman}
\email{ian.spielman@nist.gov}
\affiliation{National Institute of Standards and Technology, Gaithersburg, Maryland, 20899, USA}
\author{P.~R.~Johnson}
\affiliation{National Institute of Standards and Technology, Gaithersburg, Maryland, 20899, USA}
\author{J.~H.~Huckans}
\affiliation{National Institute of Standards and Technology, Gaithersburg, Maryland, 20899, USA}
\affiliation{University of Maryland, College Park, Maryland, 20742, USA}
\author{C.~D.~Fertig}
\thanks{Current address: University of Georgia, Athens, Georgia, 30602, USA}
\affiliation{National Institute of Standards and Technology, Gaithersburg, Maryland, 20899, USA}
\affiliation{University of Maryland, College Park, Maryland, 20742, USA}
\author{S.~L.~Rolston}
\affiliation{National Institute of Standards and Technology, Gaithersburg, Maryland, 20899, USA}
\affiliation{University of Maryland, College Park, Maryland, 20742, USA}
\author{W.~D.~Phillips}
\affiliation{National Institute of Standards and Technology, Gaithersburg, Maryland, 20899, USA}
\affiliation{University of Maryland, College Park, Maryland, 20742, USA}
\author{J.~V.~Porto}
\affiliation{National Institute of Standards and Technology, Gaithersburg, Maryland, 20899, USA}

\date{\today}

\begin{abstract}
We separate a Bose-Einstein condensate into an array of 2D sheets using a 1D optical lattice, and then excite quantized vibrational motion in the direction normal to the sheets.  Collisions between atoms induce vibrational de-excitation, transferring the large excitation energy into back-to-back outgoing atoms, imaged as rings in the 2D plane.  The ring diameters correspond to vibrational energy level differences, and edge-on imaging allows identification of the final vibrational states.  Time dependence of these data provides a nearly complete characterization of the decay process including the energies, populations, and lifetimes of the lowest two excited vibrational levels.  The measured decay rates represent a suppression of collisional de-excitation due to the reduced dimensionality, a matter wave analog to inhibited spontaneous emission.
\end{abstract}


\maketitle

Most quasi-2D quantum systems have been realized with electrons in semiconductors, where a 1D potential confines the electrons to the lowest quantized vibrational states in one direction, while leaving them essentially free in the remaining two.  Recently it has become possible to confine degenerate atomic Bose gases to 2D~\cite{Orzel2001,Gorlitz2001} and investigate vibrational excitations in the tightly confined direction~\cite{Denschlag2002}.  Trapped 2D atomic gases provide experimental opportunities unavailable in electron systems.  For example, unlike semiconductors, the atomic system is nearly defect-free.  Further, the dynamic control of the confining potential, coupled with an ability to image the atoms, enables the direct detection of the excited state population and the momentum distribution.  Quantized vibrational states are an ingredient in proposals to realize exotic states of matter, such as striped or super-solid phases~\cite{Schmid2004,Scarloa2005,Isacsson2005}, and are possible motional qubit states for quantum computation~\cite{Charron2002,Eckert2002,Porto2003}.  Stronger confinement (beyond that described herein) can also change the nature of collisions~\cite{Olshanii1998,Petrov2001,Bouchoule2002}.  

\begin{figure}[t!]
\begin{center}
\includegraphics[width=3.375in]{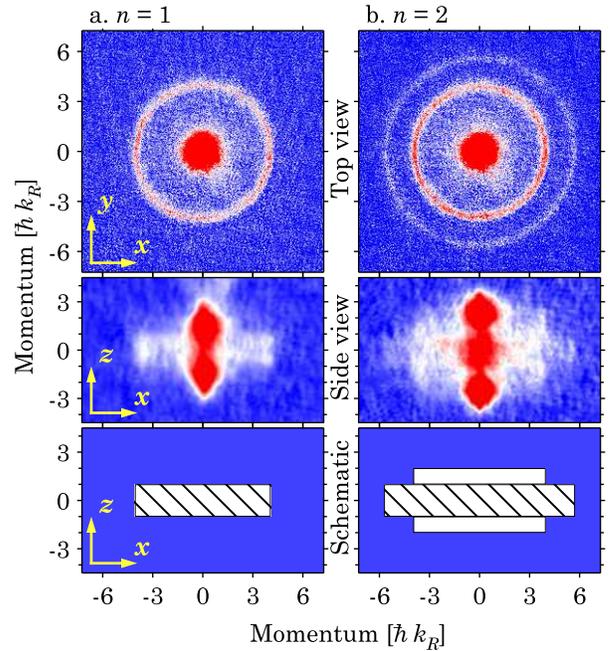}
\end{center}
\caption{Absorption images of vibrationally excited atoms after a $1\ms$ decay and subsequent TOF, with 60\% initial population in the a) $n=1$ and b) $n=2$ vibrational levels.  In the top views, outgoing rings ($t_{\rm TOF}=7.1\ms$) correspond to different in-plane energies imparted to the atoms from various decay channels.  Viewed from the side ($t_{\rm TOF}=13.1\ms$), the rings appear as rectangles, where the vertical momentum distribution identifies the final vibrational states.  The schematic of the side view shows the expected distributions from the dominant decay processes to $n=1$ (white rectangles) and $n=0$ (hatched rectangles).}
\label{rawdata}
\end{figure}

Here we study the vibrational relaxation of a quasi-2D degenerate Bose gas where quantized motion in the tightly confined direction plays a role analogous to an internal degree of freedom.  We transfer a large fraction of atoms into excited vibrational states, creating highly non-equilibrium atom populations.  In this system, atom-atom collisions provide the only significant relaxation mechanism, transferring ``internal'' energy to 2D kinetic energy.  We directly observe atom populations as outgoing rings (Fig. \ref{rawdata}) representing distinct decay channels.  The excited state lifetimes are enhanced due to the reduction in the density of final scattering states relative to scattering in an unconfined 3D gas.

We create independent 2D sheets (or pancakes) of atoms by applying a deep 1D optical lattice to a 3D Bose-Einstein condensate (BEC).  In the tight direction ($\hat z$) the system is well-described by a single-particle 1D \Schrodinger equation, yielding discrete vibrational levels labeled by an index $n$.  Atoms, Raman-excited from the ground state $n=0$ to $n=1$ or $2$, collide and decay.  By imaging after a time-of-flight $t_{\rm TOF}$ we identify the momentum and population of atoms in the various final vibrational states.  We extract excited state lifetimes from time sequences of these single-shot vibrational spectra.

We produce a magnetically trapped $\Rb87$ BEC with up to $2.5\times10^5$ atoms in the $\left|F=1,m_F=-1\right>$ state~\cite{Tolra2004}.  The BEC is separated into a stack of about $80$ pancakes by an optical lattice with period $d=410(1)\nm$ \footnote{All uncertainties herein reflect the uncorrelated combination of single-sigma statistical and systematic uncertainties.  Shot-to-shot fluctuations are the largest source of uncertainty in lattice depth and atom number.} vertically aligned along $\hat z$~\footnote{The nearly counter-propagating lattice beams are generated by a Ti:sapphire laser tuned to $\lambda = 809.5\nm$, intersecting at $\theta=162(1)^\circ$.}.  When in the ground vibrational state, the largest pancake has $N\approx4.6(10)\times10^3$ atoms with a 2D chemical potential $\mu_{\rm 2D} = 1.6(2)\kHz$; the resulting peak 2D and 3D densities are $2.2(2)\times10^9\cm^{-2}$ and $2.9(3)\times10^{14}\cm^{-3}$, respectively.  The 2D Thomas-Fermi radii are $R_x = 11.0(6)\micron$ and $R_y = 12.0(7)\micron$.  In the combined magnetic and optical potential, the in-plane oscillation frequencies are $\omega_x/2\pi=55(1)\Hz$ and $\omega_y/2\pi=50(1)\Hz$.  The lattice is raised in $200\ms$ with an exponentially increasing ramp ($50\ms$ time constant).  This timescale is chosen to be adiabatic with respect to mean-field interactions and vibrational excitations.  By pulsing the lattice and observing the resulting atom diffraction~\cite{Ovchinnikov1998}, we measure a lattice depth of $s=77(4)$, expressed in units of $\Er=h^2/8 m d^2=h\times3.42(2)\kHz$.  For a single well of the deep sinusoidal potential, the energy spacings are $E_{n+1}-E_n \simeq\Er[2\sqrt{s}-(n+1)]$ (these energies, which include the lowest order anharmonic correction, are shifted at the 1\% level by the inter-atomic interaction).  The harmonic frequency is $\omega_z/2\pi=2\Er\sqrt{s}/h=60(1)\kHz$.

Raman transitions between vibrational levels~\cite{Perrin1998} are driven by a pair of laser beams.  The nearly counter-propagating Raman beams are oriented approximately along $\hat z$~\footnote{The Raman beams have a $1/e^2$ radius $w_0=1.2\mm$ and are canted from $\hat z$, leading to a small momentum kick of $0.06\kr$ in the 2D plane.}, are detuned $82\GHz$ below the $\Rb87$ D2 transition, and have a relative detuning $\delta$ ranging from $50$ to $120\kHz$.  A $1\ms$ Raman pulse excites a fraction $f_n$ of the atom population to either $n=1$ or $n=2$.

The anharmonicity of the potential allows us to selectively transfer populations between desired vibrational levels, provided the pulse duration $t_p$ is long enough that its Fourier spread resolves the $\gtrsim\Er/h$ difference from unwanted transitions.  $t_p$ must also be shorter than the vibrational lifetime $\tau$.  For our experiment (where $\tau \gtrsim t_p\gtrsim\hbar/\Er$), we find that by detuning below Raman resonance and chirping $15\kHz$ through resonance in $1\ms$, we controllably transfer up to 65\% of the atoms to either $n=1$ or $n=2$ (simulations indicate a maximum transfer of around $75\%$ in this situation) without significantly populating unwanted states ($\sim5\%$).

The vibrationally excited atoms are allowed to decay for variable hold times $t_{\rm hold}$ ranging from $10\us$ to $10\ms$;  then the lattice is turned off in $200\us$.  The magnetic trap is then turned off in $\approx300\us$, and the atom cloud expands for $t_{\rm TOF}$ before absorption imaging.  The resulting atomic distribution, which constitutes an average over the $\sim80$ pancakes, is imaged along one of two axes: perpendicular to the pancakes (top view, along $\hat z$) and edge-on (side view, along $\hat y$).

The decay of a pair of atoms from initial vibrational states $a$ and $b$ to final states $c$ and $d$ (denoted $a b \rightarrow c d$) can be viewed as a 2D two-body inelastic collision process, where vibrational states in the confined direction play the role of internal states.  The release of vibrational energy leads to back-to-back atom pairs with large momenta in the 2D plane.  Figure \ref{rawdata} shows example TOF images from which we can infer the final momentum distribution.  Image coordinates are in units of recoil momentum, $\hbar\kr=\pi \hbar/ d$~\footnote{After TOF the atom distribution maps final position to initial momentum.  The TOF used in the computation of momentum is increased by $0.5\ms$, accounting for the $1\ms$ window during which the atoms decay.}, and the corresponding recoil velocity is $\hbar\kr/m=5.6\mms$.  Each image pair (Figs. \ref{rawdata}a and \ref{rawdata}b) contains nearly full spectral information of an excited state and its decay paths.

\begin{table}
\begin{tabular*}
{3.1 in}
{@{\extracolsep{\fill}}cccc}\hline\hline
Ring & \ \ Decay path\ \  & $E_f-E_i$ &  $I_{c d}^{a b}$ \\
\hline
Central ring &$11\rightarrow20$	& $\Er/2$				& 0.0243 \\
&$22\rightarrow31$	& $\Er/2$						& 0.0257 \\ 
&$22\rightarrow40$	& $2\Er$						& 0.0007 \\\hline
Inner ring & $11\rightarrow00$	& $\hbar\omega_z-\Er$	& 0.2202 \\
& $22\rightarrow11$	& $\hbar\omega_z-2\Er$			& 0.1556 \\
& $22\rightarrow20$	& $\hbar\omega_z-3\Er/2$		& 0.0020 \\
& $20\rightarrow00$	& $\hbar\omega_z-3\Er/2$		& 0.1128 \\
\hline
Outer ring & $22\rightarrow00$	& $2\hbar\omega_z-3\Er$		& 0.1075 \\\hline
\end{tabular*}
\caption{Release energies (including the lowest order anharmonic corrections) and matrix elements for the experimentally relevant decay paths and lattice depth $s=77$ (sorted in order of increasing decay energy).  The energies denote the per atom difference between initial and final vibrational energies.}
\label{numbers}
\end{table}

The decay paths allowed by parity and conservation of energy are shown in Table \ref{numbers}. For atoms excited to $n=1$, only two decay paths contribute.  The high-energy process, $11\rightarrow00$, gives rise to the ring in Fig. \ref{rawdata}a (top view), while the ring from the low-energy $11\rightarrow20$ process is not resolved from the central cloud.  Each ring's radius corresponds to the atomic in-plane velocity and hence the per atom difference between initial and final vibrational energies.  In Fig. \ref{rawdata}b, atoms excited to $n=2$ decay through several channels.  The processes $22\rightarrow11,$ $20\rightarrow00$, and $22\rightarrow20$ all contribute to the inner ring of Fig. \ref{rawdata}b.  Each process imparts an energy of about $2\Er\sqrt{s}$ per atom; the differences in energy of $\Er/2$ are small compared to the ring's width and are unresolved in the image.  The outer ring results solely from the $22\rightarrow00$ decay process.

\begin{figure}[t!]
\begin{center}
\includegraphics[width=3.375in]{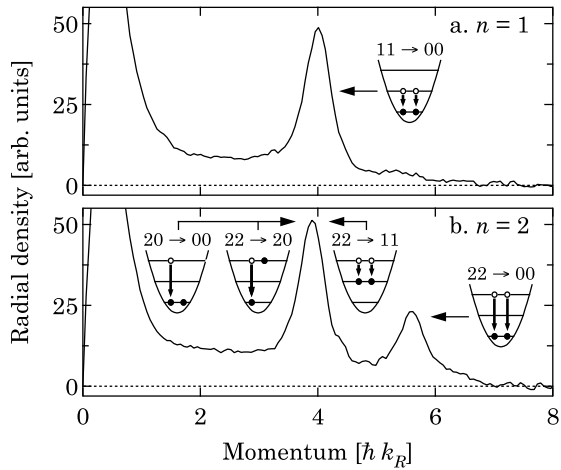}
\end{center}
\caption{Angular integrals of top view data.  a) From Fig. \ref{rawdata}a, the single ring gives rise to a peak located at $k=4.0(1)\ \kr$.  b) The two rings in Fig. \ref{rawdata}b yield two peaks at $k=3.9(1)\ \kr$ and $5.6(2)\ \kr$.  The diagrams schematically illustrate the decay channels which contribute to each peak.}
\label{angint}
\end{figure}

Figures \ref{angint}a and \ref{angint}b show the radial density of the atomic cloud after an angular integration of the data in Figs. \ref{rawdata}a and \ref{rawdata}b, respectively.  We extract the vibrational energy spacings by fitting the data to Lorentzians.  We attribute the atom background between the peaks to secondary scattering involving outgoing atoms and include it in the fits as a linearly sloping baseline.  For atoms initially in $n=1$ (Fig. \ref{angint}a), the measured energy is $h\times55(2)\kHz$, in accord with the expected value for a $77(4)\ \Er$ lattice: $E_1-E_0=h\times56.5(15)\kHz$.  For the $n=2$ data (Fig. \ref{angint}b) we measure the energies of the outer and inner rings to be $h\times107(5)\kHz$ and $h\times52(2)\kHz$, respectively.  The energies agree with the calculated values of $E_2-E_0=h\times110(3)\kHz$ and $E_2-{\bar E_1} = h\times53.8(15)\kHz$ ($\bar E_1$ is a weighted average using the matrix elements in Table \ref{numbers}).  The branching ratio between the inner and outer rings estimated from Fig. \ref{angint} (ignoring the atom background) is $2.1(3)$.  The value calculated from the matrix elements in Table \ref{numbers} is $2.9(3)$ for an excited fraction $f_2=0.60(5)$.

The side view images in Figs. \ref{rawdata}a and \ref{rawdata}b complement the top view images by identifying the final vibrational states.  The $200\us$ lattice turn-off is adiabatic with respect to the $55\kHz$ vibrational frequency.  As a result, the turn-off procedure maps quasi-momentum states in the 1D lattice to the corresponding free particle momentum states~\cite{Denschlag2002}.  For example, atoms in the $n=1$ vibrational state reside in the second Brillouin zone (BZ), and are mapped to a continuum of momentum states $k_z$ where $\kr<|k_z|<2\kr$~\footnote{The $200\ms$ turn-on procedure is adiabatic with respect to all time scales except inter-pancake tunneling.  Such a system fills the relevant BZ due to the essentially random phases that have developed in different pancakes via spatial variations in mean field shifts and trapping potentials.}.

Figure \ref{rawdata}a shows atoms which were initially excited into $n=1$.  In the side view, the atom cloud's extent along $\hat z$ reflects the mapping of quasi-momentum to momentum, and the extent in the horizontal direction reflects the final in-plane momentum.  The ring in the top view appears as a rectangle in the side view (schematically illustrated by the hatched rectangle in Fig. \ref{rawdata}a).  The dense, vertically aligned double-lobed structure at the center of the side view image is largely due to atoms which have not decayed and remain in the $n=1$ state (second BZ).  Fig. \ref{rawdata}b depicts atoms initially in $n=2$.  The process $22\rightarrow00$ gives rise to atoms in the first BZ (hatched rectangle in Fig. \ref{rawdata}b), while $22\rightarrow11$ leads to atoms in the second BZ (white rectangles in Fig. \ref{rawdata}b).  Note that the vertical central structure is taller since it contains atoms remaining in $n=2$ (third BZ).

We calculate short-time two-body branching ratios and decay rates in a single pancake using Fermi's golden rule.  Due to the extreme anisotropy of our potential the initial condensate wavefunctions can be approximated as a product of single particle wavefunctions, $\Psi_i(x,y,z)\approx\psi(x,y)\varphi_{n_i}(z)$.  Here $\psi(x,y)$ satisfies an effective 2D Gross-Pitaevskii equation~\cite{Petrov2001} for $n=0$ atoms in the Thomas-Fermi limit; we assume $\psi(x,y)$ remains unchanged during the short duration of the experiment.  $\varphi_{n_i}(z)$ solves the \Schrodinger equation for the 1D lattice potential and are nearly harmonic oscillator wavefunctions with an extent $\sigma_z=\sqrt{\hbar/m\omega_{z}}$.  The final states $\Psi_f(x,y,z)=\exp[-i (k_x x + k_y y)]\varphi_{n_f}(z)$ are free particles in $x$ and $y$ which is justified since $\omega_z\gg\omega_{x,y}$.  The rate for the scattering process $ab\rightarrow cd$ is $\Gamma_{c d}^{a b} N_a N_b$, with atom populations $N_a$ and $N_b$, and
\begin{equation}
\Gamma_{c d}^{a b}=2\pi a_s^2\omega_z I_{c d}
^{a b}\int {\rm d}x {\rm d}y\left|\psi(x,y)\right|^{4},
\end{equation}
where $I_{c d}^{a b}=2\pi\left|\sigma_{z}\int {\rm d}z\varphi_{a} (z) \varphi_{b}(z) \varphi_{c}(z)  \varphi_{d}(z)\right|^{2}$ is dimensionless (see Table \ref{numbers}), and $a_s=5.3\nm$ is the s-wave scattering length~\cite{vanKempen2002}\footnote{The 3D scattering is largely unaffected by confinement since $a_s=5.3\nm$ is much smaller than $\sigma_z\simeq44\nm$~\cite{Petrov2001}.}.  Parity considerations make $I_{c d}^{a b}=0$ when $a+b+c+d$ is odd.  The total rate is a sum over energetically allowed final states $c$ and $d$.

\begin{figure}[b]
\begin{center}
\includegraphics[width=3.375in]{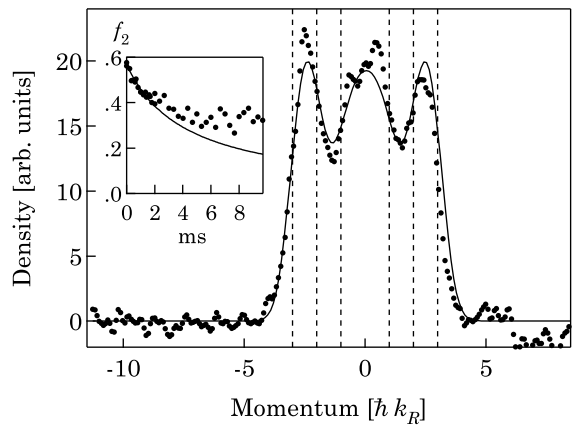}
\end{center}
\caption{Cross-sectional integral of Fig. \ref{rawdata}b (side-view).  The vertical dashed lines delineate the lowest 3 BZs.  For the specific example here, the fit (solid curve) gives fractions $f_0 = 0.37$, $f_1 = 0.18$, and $f_2 = 0.45$ in each of the BZs.  Inset: $t_{\rm hold}$ dependence of $n=2$ population for a $60\%$ initial fraction, showing the fit to the first $2\ms$.}
\label{lifetime}
\end{figure}

For atoms excited to $n=1$, the excited fraction $f_1(t)$ is governed by:
\begin{eqnarray*}
\dot{f}_1 &=&-2 N\left[\Gamma_{00}^{11} + 2\Gamma_{20}^{11} \right]f_1^2~.
\end{eqnarray*}
For atoms excited to $n=2$, the fraction $f_2(t)$ is governed by the coupled equations:
\begin{eqnarray*}
\dot{f}_2 & = & -2 N [(\Gamma_{00}^{22}+ \Gamma_{11}^{22} + 2\Gamma_{31}^{22}) f_2 + \Gamma_{00}^{20} f_0] f_2\\
\dot{f}_0 & = & -2 N \Gamma_{00}^{20} f_2 f_0~. 
\end{eqnarray*}
These equations ignore small contributions from $\Gamma_{20}^{22}$ and $\Gamma_{40}^{22}$ and assume that once atoms decay, they leave the cloud and do not re-scatter.  Parity considerations forbid the transition $10\rightarrow00$ but not $20\rightarrow00$, so as $f_1\rightarrow0$ the $n=1$ decay rate per atom $\dot{f}_1/f_1\rightarrow0$, but as $f_2\rightarrow0$ the corresponding $n=2$ rate is non-zero.

The excited-state population as a function of $t_{\rm hold}$ is found from a series of side-view images like those in Fig. \ref{rawdata}.  Integrating over $x$ gives the atomic distribution along $\hat z$; Fig. \ref{lifetime} is the $z$ distribution corresponding to Fig. \ref{rawdata}b ($t_{\rm hold}=1\ms$).  The dashed lines delineate the BZs.  In this example, atoms are predominantly in $n=0$ and $n=2$. To extract populations, we fit the data to a distribution, flat within each BZ, convolved with a Gaussian (solid curve in Fig. \ref{lifetime}).  The width of the Gaussian is fixed by applying this model to data in which only the ground state (first BZ) was occupied.  Repeating the fitting process for different $t_{\rm hold}$ yields the fractional population as a function of time (inset to Fig. \ref{lifetime}).  We extract rates by fitting the first $2\ms$ (where we anticipate our model to be valid) to the expected solutions $f_1(t)$ and $f_2(t)$~\footnote{To achieve better signal to noise, we fit the first $5\ms$ for the $n=1$, $f_1\approx0.2$ data point.}.  

\begin{figure}[t!]
\begin{center}
\includegraphics[width=3.375in]{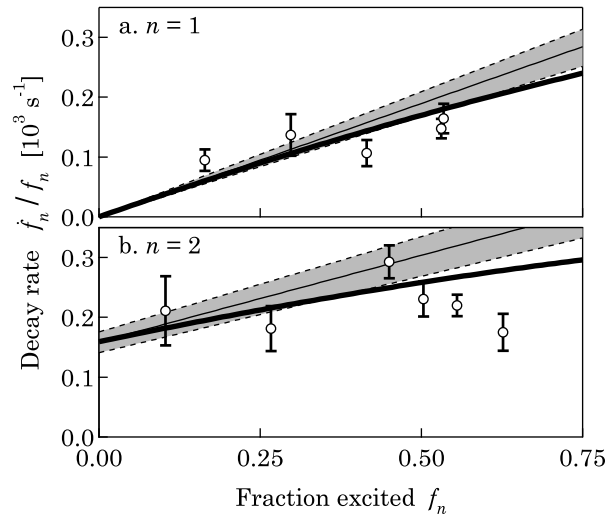}
\end{center}
\caption{Initial decay rates.  a) initial population in $n=1$.  b) initial population in $n=2$.  The thin lines represent the result of our model with no adjustable parameters, integrated over the distribution of pancakes expected from Thomas-Fermi approximation.  The grey regions reflect the variation in this model due to experimental number uncertainty.  The thick solid lines include the estimated effect of secondary scattering, using a 75\% probability of a second event.}
\label{results}
\end{figure}

The resulting rates per atom at $t=0$ are shown in Fig. \ref{results}.  For comparison we plot the predicted rates at $t=0$, where the grey band reflects the experimental atom number uncertainty. The agreement, with no free parameters, is good even though we neglected significant corrections from secondary collisions.  The presence of atoms not on the rings (Fig. \ref{rawdata} and \ref{angint}) indicates that secondary scattering is important.  An analysis of Fig. \ref{angint} suggests that around 75\% of outgoing atoms re-scatter, consistent with our theoretical estimates.  These processes decrease the atomic density, and hence reduce the overall decay rate.  (In separate experiments with lower density thermal samples, we indeed observe significantly lower rates: below $40\ {\rm s}^{-1}$.)  The effect should be more significant for large initial excited fractions, a trend consistent with the data.  The solid lines in Fig. \ref{results} show the estimated effect of secondary scattering from outgoing atoms which decayed during the $1\ms$ Raman pulse.

In conclusion, we measured the collision-induced transition rates between different vibrational states in a deep 1D optical lattice using a new, single-shot, spectrographic technique.  The reduced dimensionality of the final states suppresses the total scattering rate in a manner similar to the suppression of spontaneous emission in a planar cavity~\cite{Hulet1985}.  In a separate calculation, we find that scattering into an unrestricted 3D space exceeds our observed rates by a factor between 2.5 and 3.5 (the variation stems from different possible assumptions in the calculation).  Additional suppression of the rates would be achieved by further modifying the final density of states~\cite{Isacsson2005}, for example by confining the atoms into 1D tubes~\cite{Tolra2004}, or in analogy with photonic band-gap materials, by applying an additional in-plane lattice to open suitably placed band-gaps.  The long lifetimes are expected to be useful in the context of producing correlated atomic systems.

We thank E. Tiesinga, P. Naidon, and C. J. Williams for useful conversations.  This work was supported in part by ARDA, ONR, and NASA.  P.R.J. acknowledges support from the DCI postdoctoral research program.


\end{document}